# A Chiral Paramagnetic Skyrmion-like Phase in MnSi

C. Pappas,[1,2] E. Lelièvre-Berna,[3] P. Falus,[3] P. M. Bentley,[3] E. Moskvin,[1,4] S. Grigoriev,[4] P. Fouquet,[3] and B. Farago[3]

[1]*Helmholtz Zentrum Berlin für Materialien und Energie, Glienickerstr. 100, 14109 Berlin, Germany*[*]
[2]*Faculty of Applied Sciences, Delft University of Technology, Mekelweg 15, 2629 JB Delft, The Netherlands*
[3]*Institut Laue-Langevin, 6 rue Jules Horowitz, 38042 Grenoble, France*
[4]*Petersburg Nuclear Physics Institute, 188300 Gatchina, Leningrad District, Russia*

(Dated: April 20, 2009)

We present a comprehensive study of chiral fluctuations in the reference helimagnet MnSi by polarized neutron scattering and Neutron Spin Echo spectroscopy, which reveals the existence of a completely left-handed and dynamically disordered phase. This phase may be identified as a spontaneous skyrmion phase: it appears in a limited temperature range just above the helical transition $T_C$ and coexists with the helical phase at $T_C$.

Chirality is ubiquitous in nature and of fundamental importance both on the microscopic level and in our everyday life. The break of symmetry between right and left manifests itself in parity violation, governs biological structures such as DNA and can also be experienced in the organisation of our own body. In magnetism, chirality is evident in solitons [1], systems with geometric frustration [2] and metallic systems with non-centro-symmetric lattice structures, where the resulting anti-symmetric Dzyaloshinski-Moriya (DM) interactions [3, 4] introduce a parity breaking term in the Hamiltonian [5]. The DM term has the form $\vec{M} \times (\vec{\nabla} \times \vec{M})$ and is more than a perturbation giving rise to the peculiar canted magnetic arrangements found in high temperature superconductors [6] or the cycloid spin structures in multiferroics [7, 8]. In the non-centrosymmetric weak itinerant-electron ferromagnet MnSi, DM induced chirality comes in close interplay with Fermi liquid behavior and quantum fluctuations [9]. The Hamiltonian of MnSi comprises three hierarchically ordered magnetic interaction terms with well separated energy scales [10], which allow to distinguish between different contributions. The strongest ferromagnetic exchange interaction aligns the spins, the weaker chiral Dzyaloshinski-Moriya (DM) term twists them into a helix and the weakest Anisotropic Exchange (AE) or crystal field term pins the helix propagation vector $\vec{\tau}$ along the $\langle 111 \rangle$ crystallographic directions. The helical order appears below $T_C \approx 29$ K. It is a left-handed helix with a period of $\ell \sim 175$ Å ($\tau \approx 0.036 \text{Å}^{-1}$) and all magnetic moments perpendicular to the helix vector [11].

In this letter we concentrate on the chiral correlated paramagnetic or spin liquid phase of MnSi just above $T_C$, where intense diffuse neutron scattering spreads homogeneously over the surface of a sphere with radius $\tau$. This unusual feature emerges as a ring on the two-dimensional small angle neutron scattering patterns and the rings reduce to half-moons if the beam is polarized. This is illustrated by figure 1, which reproduces spectra from [12]. Numerous theoretical studies were devoted to explain this phase invoking possibilities such as unpinned helical order [12, 13] or condensation of chiral order parameters [14]. Recent local mean-field calculations assuming the hierarchical hamiltonian of MnSi show that the helical phase is preceded by a disordered phase with skyrmion-like short range order similar to the partial order in liquid crystals [15], which sets in at $T_{C'} \approx T_C + 1K$ (see supplementary information of [15]). Skyrmions are solutions of the non-linear field theory introduced by Skyrme [16] to bridge the gap between waves and particles in the wave-particle duality concept. Skyrmionic, vortex-like, states and lattices were first predicted in non-centrosymmetric magnetic systems by A. N. Bogdanov [17]. In MnSi a skyrmion lattice was recently identified under magnetic field [18]. At zero magnetic field skyrmions would result from a vortex-like short range order of the isotropically twisted spin liquid phase (see insert on Figure 2b).

Spin correlations and fluctuations of MnSi have been extensively studied by triple axis spectroscopy (TAS) at Brookhaven in the mid-80s [19] and more recently with polarized neutrons [20]. These studies, however, did not have the resolution needed to see fluctuations close to the helical transition. We opted for neutron spin echo spectroscopy (NSE) [21], which reaches the highest energy resolution ($<1$ neV) among all neutron scattering techniques. We also combined NSE with spherical neutron polarimetry (SNP) [22] to the polarimetric NSE technique [23] for a rigorous analysis of chiral fluctuations.

The starting point of our study is the formalism devel-oped almost simultaneously by Blume [24] and Maleyev [25] in the early '60s. The magnetic scattering cross-section for thermal neutrons may be written as:

$$\sigma = \vec{M}_\perp \cdot \vec{M}_\perp^* + \vec{P}' \cdot \Im(\vec{M}_\perp \times \vec{M}_\perp^*) \quad (1)$$

where $\vec{P}'$ is the polarization vector of the incident beam and the magnetic interaction vector $\vec{M}_\perp$ is the projection of the magnetic structure factor onto a plane perpendicular to the scattering vector $\vec{Q}$. The (imaginary) vector product is the chiral term characteristic of the spiral structure.

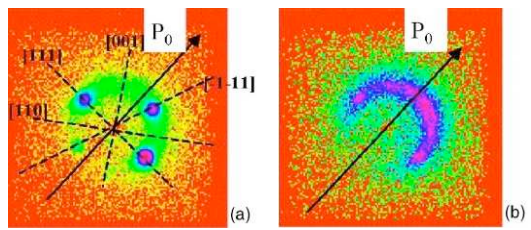

FIG. 1: Polarized neutron small angle scattering patterns measured by [12] (a) slightly below ($T_C$-0.1 K) and (b) above ($T_C$+0.2 K) the helimagnetic transition of MnSi. The vectors $P_0$ mark the polarization of the incoming neutron beam.

In MnSi the helical arrangent of the magnetic moments is described by two orthogonal vectors of the same amplitude, which are perpendicular to the helix vector $\vec{\tau}$. At the helical Bragg peaks $\vec{\tau}_{111}$ the sample orientation is such that for a left handed helix the cross-section is maximum for $\vec{P}' = \hat{Q}$ and completely extinct for $\vec{P}' = -\hat{Q}$ [24]. This extinction is responsible for the characteristic half-moon small angle scattering patterns seen with polarized neutrons above $T_C$ and gives a very precise determination of the weight of the dominant chiral domain $\eta = |\vec{M}_\perp \times \vec{M}_\perp^*|/(\vec{M}_\perp \cdot \vec{M}_\perp^*)$. $\eta$ can also be determined from the polarization vector of the scattered neutron beam $\vec{P}$:

$$\vec{P}\sigma = -\vec{P}'(\vec{M}_\perp \cdot \vec{M}_\perp^*) + 2\Re((\vec{P}' \cdot \vec{M}_\perp^*)\vec{M}_\perp) \ldots$$
$$\ldots -\Im(\vec{M}_\perp \times \vec{M}_\perp^*) \quad (2)$$

The first two terms lead to the conventional $\pi$ rotation of the polarization vector around $\vec{M}_\perp$. The chiral term creates a polarization (anti)parallel to $\hat{Q}$ independently from the incoming beam polarization. Complete knowledge of Eq. 2 implies complete knowledge and control of $\vec{P}'$ and $\vec{P}$ independently from each other. This is achieved with spherical neutron polarimetry using Cryopad, which brings $\vec{P}'$ and $\vec{P}$ either along $\hat{x} = \hat{Q}$, $\hat{z}$ (perpendicular to the scattering plane) or $\hat{y}$, which completes the right-handed cartesian set [26]. This procedure determines the polarization matrix:

$$\mathbb{P}_{\alpha,\beta} = \frac{P_{\alpha,\beta}}{|\vec{P}'|} = \frac{\widetilde{P}_{\alpha,\beta}P'_\alpha + P^\dagger_\beta}{|\vec{P}'|} \text{ with } (\alpha,\beta) \in \{x,y,z\} \quad (3)$$

where $\widetilde{P}$ is the polarization tensor of the non-chiral terms and $\vec{P}^\dagger$ the polarization created by the chiral term. At $\vec{\tau}_{111}$, $\hat{x} \parallel \vec{\tau}$ and only the first row of the matrix is non-zero. The ideal matrix is given in Table I, where $\zeta$ determines the chirality of the helix: $\zeta = +1$ for right and $\zeta = -1$ for left handed chirality. $\eta$ measures the fraction of the dominant chiral domain. $\eta = 1$ for a single domain and $\eta = 0$ for a disordered state or equally populated chiral domains.

The polarimetric measurements were done with Cryopad on the NSE spectrometer IN15 of the ILL. With the polarimetric NSE configuration it was possible to determine the matrix $\mathbb{P}_{\alpha,\beta}$ and at the same time access the intermediate scattering function I(Q,t) of each non-zero term. These measurements were complemented by "conventional" NSE measurements on IN11 (ILL). The measurements were done at λ=9 Å (IN15) and 5.5 Å (IN11) with $\Delta\lambda/\lambda$=15% FWHM and a $Q$-resolution of $\sim 0.005\,\text{Å}^{-1}$ FWHM. Most of the measurements were carried out at $\vec{\tau}_{111}$. Additional spectra were recorded on IN11 around $\vec{\tau}_{110}$ (the position of $\vec{\tau}_{110}$ is shown on the SANS patterns of figure 1). It was also possible to extract precise information on the $q = Q - \tau$ dependence of the relaxation by analyzing slices of the detector at constant distance from the 000 point. The MnSi single crystal was the same as in previous experiments [12] with the B20 lattice structure ($P\,2_1\,3$) and a lattice constant of 4.558 Å.

TABLE I: Ideal and measured polarization matrix $\mathbb{P}_{\alpha,\beta}$ at $\vec{\tau}_{111}$. For details see text.

| $\beta \backslash \alpha$ | ideal | | | T = 25 K | | |
|---|---|---|---|---|---|---|
| | $x$ | $y$ | $z$ | $x$ | $y$ | $z$ |
| $x$ | -1 | $\eta\zeta$ | $\eta\zeta$ | -0.99(1) | -0.99(1) | -0.99(1) |
| $y$ | 0 | 0 | 0 | 0.05(2) | 0.05(2) | 0.05(2) |
| $z$ | 0 | 0 | 0 | -0.01(2) | -0.01(2) | -0.01(2) |

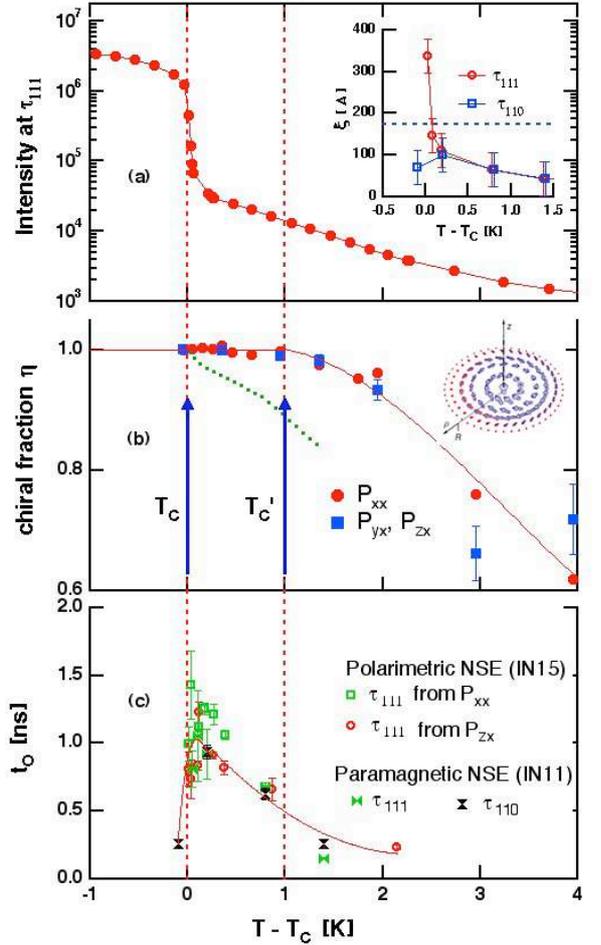

FIG. 2: Intensity at the position of the $\vec{\tau}_{111}$ helical peak (a), chiral fraction of the dominant left handed domain (b) and characteristic time $t_0$ of the fluctuations at $\vec{\tau}_{111}$ and $\vec{\tau}_{110}$ (c). In (b) the red circles are deduced from the extinction ($\mathbb{P}_{x,x}$ eq. 1), the blue squares from the non-diagonal terms $\mathbb{P}_{y,x}$ and $\mathbb{P}_{z,x}$ and the green dotted line is calculated from $\xi_{111}$ and eq. 4. The error bars are calculated from the counting rates and do not include systematic errors. For the sake of comparison between different cryogenic setups the data are plotted against $T - T_C$. The lines are guides for the eyes.

The transition temperature $T_C = 29.05 \pm 0.05$ K is defined from the sudden increase by more than one order of magnitude of the neutron intensity at $\vec{\tau}_{111}$ (figure 2a), which confirms the weak first order transition with strong second-order features seen on specific heat [27].

The Ornstein-Zernike analysis of the $q$ dependence of the intensity around $\vec{\tau}_{111}$ and $\vec{\tau}_{110}$ leads to correlation lengths $\xi_{111}$ and $\xi_{110}$, which are very close to previous results [12] and are shown on the insert of figure 2a. These quantify the extent of correlations along the helix vector $\vec{\tau}$. $\xi_{110}$ stays significantly smaller than the pitch of the helix and shows a maximum slightly above $T_C$, whereas $\xi_{111}$ diverges and reaches the pitch of the helix (dashed line) at $\sim T_C$+0.05 K triggering the first order transition.

The polarization matrix was measured with an incident beam polarization of 97% and the data were corrected for the background signal. At $\vec{\tau}_{111}$ the calculated and measured matrices of the helical phase at 25 K are in excellent agreement with each other (Table I) confirming the stabilisation of a single-domain left-handed helix with $\eta = 1, \zeta = -1$. When increasing the temperature, the polarization matrix $\mathbb{P}_{\alpha,\beta}$ is unaffected by the first order transition and remains unchanged up to $T_{C'} \approx T_C+1$ K. Also the extinction of the scattering when $\vec{P}' = -\hat{\tau}$ (eq. 1) is complete up to $T_{C'}$. All values of $\eta$, plotted in figure 2b, reveal the existence of a new phase between $T_C$ and $T_{C'}$. This is a completely single domain phase even though the correlations are significantly shorter than $\ell$: at $T_{C'}$ the correlation length does not exceed $\sim\ell/4$. This result contrasts with the common experimental [28] and theoretical [12] picture based on the hierarchy of the interactions, according to which MnSi behaves as a ferromagnet at distances shorter than $\ell$ and for $\xi < \ell$. In fact our experimental results show a more complex situation and the slow decrease of the chiral fraction $\eta$ displayed in figure 2c spreads over a very wide temperature range. Even at 50 K ($T_C$+20 K) where correlations do not exceed 7 Å [19], we found $\eta \sim 0.203 \pm 0.024$.

A mean field description of paramagnetic fluctuations in a ferromagnet in the presence of DM interactions was given in [12] and renders the main features of MnSi: extinction of the scattering at $\vec{P}' = -\hat{\tau}$ or intensity maximum at the sphere $Q = \tau$ (figure 1). The theory also gives a simple relation between $\eta$ and the correlation length $\xi$:

$$\eta = 2Q\tau/(Q^2 + \tau^2 + 1/\xi^2) \qquad (4)$$

From the measured $\xi_{111}$ values we calculated the dotted green line on figure 2, which is significantly lower than the experimental data. Consequently, the simple assumption of an unpinned or fluctuating helical phase is not sufficient to describe the experimental result. The high values of $\eta$ reflect a partial or short range order underlining the non-trivial character of the spin liquid phase of MnSi, in particular of the phase between $T_C$ and $T_{C'}$.

In the following we will discuss the results of our NSE measurements. NSE spectroscopy gives the normalized intermediate scattering function $I(Q,t) = S(Q,t)/S(Q,0)$, which is the time-energy Fourier transform of the scattering function S(Q,ω). The fluctuations of the single domain chiral spin liquid phase were measured with standard NSE. The fluctuations for $\vec{P}' \perp \tau_{111}$ ($\mathbb{P}_{z,x}$ or $\mathbb{P}_{y,x}$) and $\vec{P}' \| \tau_{111}$ ($\mathbb{P}_{x,x}$) at the $\vec{\tau}_{111}$ position were measured with polarimetric NSE. Figure 3 shows the intermediate scattering function for the non-diagonal term $\mathbb{P}_{z,x}$. We find exponential relaxations superimposed on an elastic contribution due to the Bragg peak, which were fitted by the function $I(Q,t) = a\exp(-t/t_0) + (1-a)$. The elastic fraction $(1-a)$ evolves from $\sim 20\%$ to 100% within 0.2 K following the fast increase of the intensity in figure 2a. The onset of the Bragg peak masks the decay of $I(Q,t)$ at $T_C$. On the other hand, at $\vec{\tau}_{110}$ there are no Bragg peaks and fluctuations may be seen also below $T_C$. The measurements were extended up to $T_C + 2$ K, where they agree with the previous TAS [19, 20] and preliminary resonance NSE results [29].

All values of $t_0$ are plotted against temperature in figure 2c. The existence of finite relaxation times at these temperatures rules out the interpretation in terms of quasi-static correlations suggested by [9]. The measured relaxation times are rather typical for correlated systems such as magnetic nanoparticles [30] or ferromagnets at $T_C$ [31]. All $t_0$ (at $\vec{\tau}_{111}$ and $\vec{\tau}_{110}$) fall on top of each other within experimental error bars: the direction of the incident polarization vector or the particular position on the sphere with radius $\tau$ have no effect on the observed dynamics. The fluctuations are governed by the correlation length at $\tau_{110}$ without any sign of slowing down at $\tau_{111}$, i.e. at the position where the Bragg peak appears. Consequently, the fluctuating chiral spin liquid phase does not develop to the helical phase. The two phases coexist close to $T_C$ and $\xi_{111}$ may be decomposed as: $\xi_{111} = \alpha\,\xi_{fluct} + (1-\alpha)\,\xi_{static}$ with $\xi_{fluct} = \xi_{110}$ reflecting the fluctuating spin liquid fraction $\alpha$ and $\xi_{static}$ the static component of the helical phase. It has been suggested that fluctuations may drive the transition in MnSi to first order [10, 32, 33]. Our present high resolution data rule out any interplay between fluctuations and the first order helical transition.

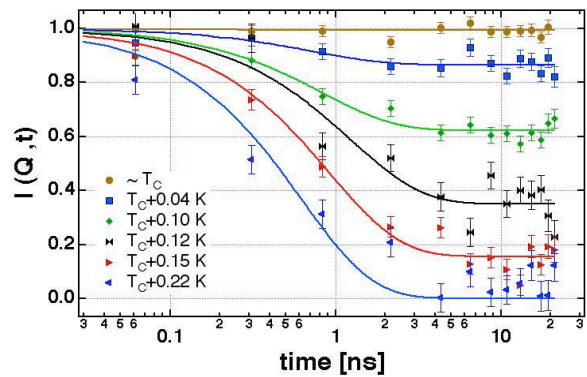

FIG. 3: Dynamic correlations at the position of one of the helical Bragg peaks $\vec{\tau}_{111}$ measured by polarimetric NSE for the $\mathbb{P}_{z,x}$ term of the polarization matrix (Table I).

Dynamic scaling relates the characteristic relaxation time to the correlation length: $t_0 \propto \xi^z$, with $z$ the dynamic exponent. Below $T_{C'}$ we found $t_0 \propto \xi_{fluct}$ leading to $z \approx 1$, an unconventional value much lower than in 3D ferromagnets ($z=5/2$) or antiferromagnets ($z=3/2$). Such a low value could reflect a locally reduced dimensionality with the spins pinned at the local helical planes.

Above $T_C$ the Bragg peaks disappear and the correlation length becomes considerably shorter than the pitch of the helix. In spite of that all spins remain pinned into the local helical plane up to $T_{C'}$. This is exactly the theoretically predicted temperature of formation of a skyrmionic ground state in MnSi [15]. It is also the temperature range of a broad maximum in the specific heat, which reflects strong short range correlations [27]. Besides specific heat, thermal expansion, sound velocity, sound absorption as well as resistivity measurements point towards a non trivial spin liquid phase [34]. Skyrmions account for the inherent topology of this single domain spin liquid phase, which gives a scattering with rotational symmetry. Skyrmionic textures are energetically favourable at short distances, which privileges structures with magnetic moments decreasing with increasing distance from the vortex center [15]. In the spin liquid phase the mean magnetic moment decreases continuously with increasing distance $d$ following the typical correlation function $\sim \exp(-d/\xi)/d$. The correlation length $\xi_{fluct} = \xi_{110}$ would, therefore, give a natural measurement of the skyrmion radius, which would not exceed $\ell/2$ at $T_C$ stabilizing the single domain spin liquid phase. That skyrmionic solutions are energetically more favourable at short lengths would explain the slow decrease of $\eta$ above $T_C$ and its finite value even at 50 K.

In summary, these first polarimetric NSE measurements provide the missing link in getting a consistent picture of the spin liquid phase of MnSi above $T_C$. High resolution neutron spin echo spectroscopy and polarimetry evidence the existence of a disordered skyrmion phase, a completely chiral single domain and strongly fluctuating new state of matter in a very narrow temperature range of $\sim 1$ K above the helical phase of MnSi.

C.P. thanks P. Böni and E.L.-B. thanks P.J. Brown for fruitful discussions. The authors acknowledge the support of the ILL technical teams in particular E. Bourgeat-Lami, C. Gomez and E. Thaveron. Special thanks go to Thomas Krist for the compact solid state polarizer, which enabled the polarimetric NSE measurements. This project was partly supported by the European Commission under the $6^{th}$ Framework Programme through the Key Action: Strengthening the European Research Area, Research Infrastructures. Contract no: RII3-CT-2003-505925."